\documentstyle[twoside,fleqn,espcrc2,epsf]{article}

\newcommand{\AmS}{{\protect\the\textfont2
  A\kern-.1667em\lower.5ex\hbox{M}\kern-.125emS}}

\title{An Effective Action for Finite Temperature QCD with Fermions}

\author{Peter N. Meisinger
        and 
        Michael C. Ogilvie\address{Department of Physics, Washington University,
        \\~One Brookings Drive, Box 1105, St. Louis, MO 63130}\thanks{This work
was supported by the U.S. Department of Energy under Contract
No. DE-FG02-91-ER40628. Additional support was provided to PNM by the U.S.
Department of Education in the form of a GANN Predoctoral Fellowship.}}

\begin{document}

\begin{abstract}
Using lattice perturbation theory at finite temperature, we compute for
staggered fermions the one-loop fermionic corrections to the spatial and
temporal plaquette couplings as well as the leading $Z_N$ symmetry breaking
coupling. Numerical and analytical considerations indicate that the finite
temperature corrections to the zero-temperature calculation of A.~Hasenfratz and
T.~DeGrand are small for small values of $\kappa = {1\over 2m_F}$, but become
significant for intermediate values of $\kappa$. The effect of these finite
temperature corrections is to ruin the agreement of the Hasenfratz-DeGrand
calculation with Monte Carlo data.  We conjecture that the finite temperature
corrections are suppressed nonperturbatively at low temperatures, resolving this
apparent disagreement. The $Z_N$ symmetry breaking coupling is small; we argue
that it may change the order of the transition while having little effect on the
critical value of $\beta$.
\end{abstract}

\maketitle

\section{INTRODUCTION}

It has been known for some time that the effects of heavy dynamical fermions can
be included in Monte Carlo simulations using a hopping parameter expansion of
the fermion determinant. This is done routinely in perturbative QED where
electron loops are included in an effective action. Recently Hasenfratz and
DeGrand \cite{HaDe94}\cite{Lat93} have performed a zero-temperature calculation
of the shift in the lattice gauge coupling constant induced by staggered
dynamical fermions and applied the result to the finite temperature phase
transition in QCD. Their result for the shift in the critical coupling, in the
form $\beta^{\rm pure}_c = \beta_c + \Delta\beta_F$, was found to hold rather
well down to small values of the fermion mass, even for $N_t = 4$; this is
particularly surprising since $\Delta\beta_F$ is calculated using lattice
perturbation theory at zero-temperature. In order to understand the effects of
finite temperature, we have calculated the one-loop fermionic corrections to the
spatial and temporal plaquette couplings, as well as the leading $Z_N$ symmetry
breaking coupling.

\section{RENORMALIZATION OF $\beta$ AT FINITE TEMPERATURE}

The one-loop finite temperature fermionic correction to the $O(A^2_\mu)$ term in
the gauge field lattice action is given by
\begin{eqnarray}
S_{\rm eff} = - g^2 \sum_{p_0} {1\over {N_t}} \int_{-\pi}^{~\pi} {d^3\vec{p}
\over {(2\pi)^3}} {\rm Tr}_c \Big\{ \tilde{A}_{\mu}(p)~~~~~&& \nonumber\\
\times~\tilde{A}_{\nu}(p) {\beta \over{2N_c}} \Bigl[
D_{\mu\nu}^{(0)}(p) + D_{\mu\nu}^{(1)}(p) - D_{\mu\nu}^{(2)} \Bigr] \Big\}&&
\end{eqnarray}
with
\begin{eqnarray}
D_{\mu\nu}^{(0)}(p) = 4 \Bigl[\delta_{\mu\nu} \sum_{\alpha}
\sin^2\Bigl({p_{\alpha} \over 2}\Bigr)&& \nonumber\\
- \sin\Bigl({p_{\mu} \over 2}\Bigr) \sin\Bigl({p_{\nu} \over 2}\Bigr)\Bigr]&&
\end{eqnarray}
\begin{eqnarray}
D_{\mu\nu}^{(1)}(p) = {1 \over 2} \sum_{k_0} {1 \over {N_t}} \int_{-\pi}^{~\pi}
{d^3\vec(k) \over {(2\pi)^3}} {\rm Tr}_d \Bigl[ R(k_{\mu})~~&&
\nonumber\\ \times~S^{-1}(k-p/2)R(k_{\nu}) S^{-1}(k+p/2) \Bigr]&&
\end{eqnarray}
\begin{eqnarray}
D_{\mu\nu}^{(2)} = {1 \over 2} \delta_{\mu\nu} \sum_{k_0} {1 \over {N_t}}
\int_{-\pi}^{~\pi} {d^3\vec(k) \over {(2\pi)^3}} {\rm Tr}_d
\Bigl[ Q(k_{\mu})&& \nonumber\\ \times~S^{-1}(k) \Bigr]&&
\end{eqnarray}
where the vertex functions are given by
\begin{equation}
R(k_{\mu}) = i\gamma_{\mu}\cos(k_{\mu}) \quad
Q(k_{\mu}) = -i\gamma_{\mu}\sin(k_{\mu})
\end{equation}
with no sum over $\mu$, and the inverse fermion propagator is given by
\begin{equation}
S(k) = {1 \over {2\kappa}} + i\gamma_{\mu}\sin(k_{\mu}).
\end{equation}

At finite temperature there are two independent tensors of order $p^2$ which are
four-dimensionally transverse\cite{KaKa85}. This manifests itself as separate
renormalizations of the spatial and temporal gauge couplings. We find that
\begin{equation}
\Delta\beta_s = - N_c \sum_{k_0} {1\over {N_t}} \int_{-\pi}^{~\pi} {d^3\vec{k}
\over {(2\pi)^3}} \Phi(k;1,2)
\end{equation}
with
\begin{eqnarray}
\Phi(k;\mu,\nu) = 32B^{-2}(k)\cos^2(k_{\mu})\cos^2(k_{\nu})&& \nonumber\\
- 4096B^{-4}(k)\sin^2(k_{\mu})\cos^2(k_{\mu})&& \nonumber\\
\times~\sin^2(k_{\nu})\cos^2(k_{\nu})&&
\end{eqnarray}
and
\begin{equation}
B(k) = {1 \over {\kappa^2}} + 4 \sum_{\alpha} \sin^2(k_{\alpha}).
\end{equation}
A different result is obtained for $\Delta\beta_t$ by replacing $\Phi(k;1,2)$
in (7) with $\Phi(k;0,1)$. Figure 1 compares $\Delta\beta$ per fermion
(spatial and temporal) vs. $\kappa$ for $N_t = 4$ and $N_t = 6$ with the zero
temperature result of Hasenfratz and DeGrand. The finite temperature values
approach the zero temperature result from below as a consequence of the
antiperiodic boundary conditions. Note that they ruin the agreement between the
zero-temperature calculation and the Monte Carlo results for intermediate to
large values of $\kappa$.

\begin{figure}[htb]
\bigskip
\bigskip
\bigskip
\bigskip
\bigskip
\bigskip
\bigskip
\bigskip
\bigskip
\bigskip
\bigskip
\bigskip
\bigskip
\bigskip
\bigskip
\bigskip
\bigskip
\includegraphics{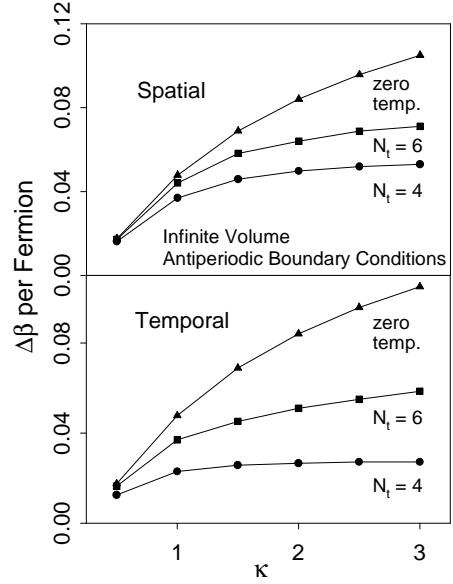}
\caption{$\Delta\beta$ per fermion (spatial and temporal) vs. $\kappa$ for
 various $N_t$.}
\label{fig:f1}
\end{figure}

The connection between the the zero and finite temperature result can be
understood more physically by transforming the sum over Matsubara frequencies
into a sum over images using the Poisson summation formula for antiperiodic
boundary conditions. The shift in the spatial plaquette coupling is then given
by
\begin{eqnarray}
\Delta\beta_s = - 2 N_c \sum_{n=0}^{\infty} {(-1)^n \over {2^{\delta_{n0}}}}
\int_{-\pi}^{~\pi} {d^4k \over {(2\pi)^4}} \Phi(k;1,2)&& \nonumber\\
\times~\cos(nN_tk_0)&&
\end{eqnarray}
with a similar result for $\Delta\beta_t$. This form has a simple physical
interpretation: the integer $n$ labels the net number of times the fermion wraps
around the lattice in the temporal direction. Numerically, the dominant
corrections to the zero-temperature result come from the first few values of
$n$, with the $n = 1$ and $n = 2$ terms accounting for more than $90\%$ of the
finite temperature correction for $\kappa \leq 2.0$.

\begin{figure}[htb]
\includegraphics{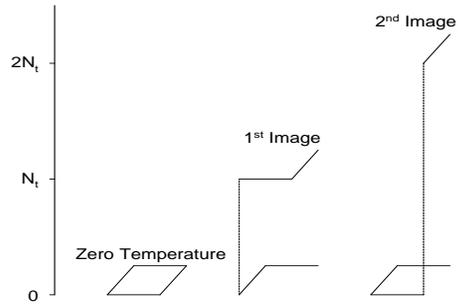}
\bigskip
\bigskip
\bigskip
\bigskip
\bigskip
\bigskip
\bigskip
\bigskip
\caption{Diagrams contibuting to finite temperature corrections for the
renormalization of $\beta$.}
\label{fig:f2}
\end{figure}

The finite temperature corrections result from the $O(A^2_\mu)$ expansion of
image diagrams such as those depicted in Figure 2. The vertical segments of the
image diagrams are powers of Polyakov loops. Given that $Z_3$ symmetry is
maintained in the confined phase of QCD, the contibutions of these terms is
negligible below $\beta_c$. Thus, the finite temperature corrections to $\beta$
may be suppressed nonperturbatively in the confined regime. In particular, near
$\beta_c$ the zero-temperature result will hold for $\Delta\beta$.

\section{$Z_N$ SYMMETRY BREAKING IN THE EFFECTIVE ACTION}

There is another set of terms induced by the fermion determinant at finite
temperature \cite{HaKaSt83}. Using the hopping parameter expansion and the
Poisson summation formula, we find an additional contribution to the effective
action:

\begin{equation}
S_{\rm eff} = \sum_{\vec{x}} \sum_{n=1}^{\infty} (-1)^{n+1} h_n(\kappa)
Re \Bigl[ {\rm Tr} P^n(\vec{x}) \Bigr]
\end{equation}
where $P(\vec{x})$ denotes a Polyakov loop and the couplings $h_n$ are given by
\begin{eqnarray}
h_n(\kappa) = -4 N_t \int_{-\pi}^{~\pi} {d^4q \over {(2\pi)^4}} {\rm ln} \Biggl[
{1 \over {4\kappa^2}}~~~~&& \nonumber\\ + \sum_{\mu} \sin^2(q_{\mu}) \Biggr]
\cos(nN_tq_0)&&
\end{eqnarray}

The maximum values of the $h_n$ are obtained when $m_F = 0$. For $N_t = 4$ we
find $h_1^{\rm max} = 0.107$, $h_2^{\rm max} = 0.00445$, and increasingly
smaller values for higher order couplings.  Including just the $h_1$ term in
the action of an otherwise pure QCD Monte Carlo simulation, we observe that this
additional source of $\Delta\beta$ is small compared to the renormalization of
the plaquette couplings discussed in the preceeding section. For example, even
$h_1 = 0.1$ at $N_t = 4$ leads to a shift in $\beta$ per fermion of $0.00325$.
This value of $h_1$ corresponds to $m_F = 0.17$ which yields a zero-temperature
predicted shift in $\beta$ per fermion of $0.104$. 

The effect of the $h_1$ term in the effective action, however, is not trivial.
We find a first-order phase transition for values of $h_1$ smaller than
approximately $0.08$. Including zero-temperature plaquette coupling
renormalization, the endpoint of this first-order phase transition line maps to
the point $(0.394, 4.68)$ in the $(m_F, \beta)$-plane for the case of sixteen
degenerate staggered fermion species. This is near the endpoint of the
first-order phase transition observed in simulations with dynamical staggered
fermions \cite{OhKi91}.

A variety of workstations were used to obtain the results presented in this
section. Typically $42000$ sweeps were made on $10^3 \times 4$ lattices.
Definitive numerics will require substantially greater computational resources.

\section{CONCLUSIONS}

Finite temperature corrections to the gauge coupling renormalization lift the
degeneracy of the spatial and temporal couplings, but the results are in
conflict with Monte Carlo data. An additional finite temperature shift in
$\beta$, due to an induced coupling to Polyakov loops, is negligible. This
coupling does, however, influence the order of the transition and may be the
most important factor in determining the end point of the first-order
deconfinement phase transition line. As we have shown, it is plausible that the
success of the zero-temperature calculation in determining the critical value
of $\beta$ is due to a nonperturbative suppression of finite temperature effects
in the low temperature regime. Specifically, the small expectation value of the
Polyakov loops at low temperatures indicates a suppression of those quark paths
which account for finite temperature corrections.

\end{document}